\documentstyle[11pt]{article} 

\begin{document} 

\title{General-Relativistic Equations of Motion 
in terms of Energy and Angular Momentum} 

\author{Yong Gwan Yi} 

\maketitle 

\begin{abstract} 

An attempt is made to describe the general-relativistic equations 
of motion for the Schwarzschild geometry in terms of the classical 
concepts of energy and angular momentum. Using the customary 
terms the geodesic equations can be viewed in a way that is 
very helpful in providing the physical meaning of the mathematical 
development. 

\end{abstract} 

\bigskip 

The general theory of relativity has led to a completely 
new picture of gravitational phenomena in geometrical terms. 
The gravitational field is represented by metric tensor, and 
the equations of free fall are geodesics. Although the geodesic 
equation gives a constant of motion corresponding to energy, 
accordingly, most textbooks introduce approaches that 
exclude serious use of energy concept [1]. However, it would 
seem desirable to use the relativistic energy in describing 
the central force problem. For many applications, the equation 
of motion containing the energy and angular momentum is the 
natural one. In order to discuss the comparison with Newton's 
theory or the transition to quantum theory, it is important that the 
description of the motion be in terms of its energy and angular 
momentum. In a certain sense, the use of relativistic energy is 
considered necessary and important. 

Let us see what can be learned from Einstein's theory of 
gravitation. In this paper, we define $a_{\mu}b^{\mu}$ as 
$a_0b^0 - {\bf a} \cdot {\bf b}$. We begin by pointing out that 
the metric of our space-time is 
$ds^2=g_{\mu\nu}dx^{\mu}dx^{\nu}$. 
The Schwarzschild expression for the metric around 
a mass $M$ is 
\begin{equation} 
c^2d\tau^2=c^2g_{00}dt^2 
-g_{rr}dr^2-r^2d\theta^2-r^2\sin^2\theta d\psi^2 
\end{equation} 
with $g_{00}=1/g_{rr}=1-2GM/c^2r$. 
We consider the central force motion in the plane $\psi=\pi/2$. 
The square of velocities will then be 
\begin{equation} 
c^2\rightarrow c^2g_{00} \quad\mbox{and}\quad v^2\rightarrow g_{rr} 
\biggl(\frac{dr}{dt}\biggr)^2+r^2\biggl(\frac{d\theta}{dt}\biggr)^2. 
\end{equation} 
In consequence of this relation the Schwarzschild metric 
can be written 
\begin{equation} 
d\tau=g_{00}^{1/2}dt\biggl(1-\frac{v^2}{c^2}\biggr)^{1/2}, 
\end{equation} 
to a first approximation. This form of equation reduces to the 
familiar equation leading to the Lorentz time dilation in the 
limit as $g_{00}$ approaches to unity. In this sense one may 
see the relation in (3) as the Schwarzschild time dilation. 

The equations of motion in the Schwarzschild field yield two 
constants of motion. One of them is given by 
\begin{equation} 
g_{00}\frac{dt}{d\tau}=\mbox{constant}, 
\end{equation} 
which corresponds to the energy of the system. The other constant 
is obtained from $r^2(d\theta/d\tau)=$ constant, and is absorbed 
immediately into the definition of the angular momentum $l$. It would 
seem at first sight that the constant in (4) is of no importance in the 
geometrical approach. However, the constant has an important 
physical significance, for it can lead to the formulation of the resulting 
relativistic mechanics in terms of the energy of a particle as in 
the case of special relativity. The relativistic equations of motion 
must be such that in the nonrelativistic limit they go over into the 
customary forms given by Newton's theory. Thus the task of 
identifying the constant is greatly facilitated by seeking the form 
which it would have in the nonrelativistic limit. In the nonrelativistic 
limit, Eq. (4) can be expanded as 
\begin{equation} 
g_{00}\frac{dt}{d\tau}\simeq \frac{1}{mc^2}\biggl(mc^2+ 
\frac 12mv^2-\frac{GMm}{r}\biggr), 
\end{equation} 
where $m$ is the mass of a particle. By comparison with Newton's 
theory, we can identify the constant with 
\begin{equation} 
g_{00}\frac{dt}{d\tau}=\frac{1}{mc^2}\bigl(mc^2+E\bigr). 
\end{equation} 
Consequently it yields the expression 
\begin{equation} 
E=mc^2g_{00}\frac{dt}{d\tau}-mc^2 
\end{equation} 
for the energy of a particle in the static isotropic 
gravitational field. 

The geodesic equations teach us a four-velocity of the form 
$g_{\mu\mu}dx^{\mu}/d\tau$. In the Schwarzschild 
metric the scalar product of two four-vectors is defined as 
$g_{\mu\nu}a^{\mu}b^{\nu}$ or 
$g^{\mu\nu}a_{\mu}b_{\nu}$. 
With this definition the square of the magnitude of the velocity 
four-vector is a constant, $c^2$. From the covariant form of 
velocity we can write the relativistic expression for momentum 
as 
\begin{equation} 
p_r=mg_{rr}\frac{dr}{d\tau} \quad\mbox{and}\quad 
p_{\theta}=mr^2\frac{d\theta}{d\tau}\equiv l. 
\end{equation} 
This is a definition of momentum which is deduced from 
the constants of motion in the Schwarzschild metric field. 
Equations (7) and (8) are the necessary relativistic 
generalizations for the energy and momentum of a particle, 
consistent with the conservation laws and the postulates 
of general relativity. 

As in the special theory of relativity, it is natural to attempt 
to identify the four equations of energy and momentum 
conservation as relations among the energy-momentum 
four-vectors. In special relativity, the connection between the 
kinetic energy $T$ and the momentum is expressed in the 
statement that the magnitude of the momentum four-vector 
is constant: 
\begin{equation} 
p_{\mu}p^{\mu}=\frac{T^2}{c^2}-p^2=m^2c^2. 
\end{equation} 
This must be generalized to provide an expression satisfying 
the general-relativity formulation. We observe that the momentum 
in (8) is proportional to the space components of the 
four-vector velocity. The time component of the four-vector 
velocity is $cg_{00}dt/d\tau$. Comparison with (7) shows 
that the energy of a particle differs from its time component 
by the rest energy $mc^2$. We are thus led to 
\begin{equation} 
E=mc^2g_{00}\frac{dt}{d\tau} 
\end{equation} 
as the covariant form of the total energy, for then $p_r$, 
$l$, and $E/c$ form a four-vector momentum. The desired 
generalization of energy-momentum equation must be 
\begin{equation} 
g^{\mu\mu}p_{\mu}p_{\mu}=\frac{E^2}{c^2g_{00}} 
-\frac{p_r^2}{g_{rr}}-\frac{l^2}{r^2}=m^2c^2. 
\end{equation} 
It should be noted that the gravitational potential lends 
itself to incorporation in the metric of space-time geometrization, 
so the potential energy is absorbed automatically into the path length 
of a particle and its motion therein. In general relativity, therefore, 
kinetic energy and potential energy individually become meaningless; 
only the total energy of a particle is significant. 

We can now proceed to the relativistic equation for the orbit of 
a planet. We can still talk in terms of the system energy and 
the system angular momentum. For comparison with Newton's theory, 
it is preferable to define the energy $E$ as in (7), which would bring 
$E$ in line with the nonrelativistic value. The Schwarzschild metric 
in (1) can now be expressed in terms of two constants of motion 
$E$ and $l$ as 
\begin{equation} 
\frac{(mc^2+E)^2}{c^2g_{00}}-m^2g_{rr} 
\biggl(\frac{dr}{d\tau}\biggr)^2-\frac{l^2}{r^2}=m^2c^2. 
\end{equation} 
This form of the equation of motion can also be obtained from 
a combination of the differential equations of geodesics [2].
Most often we are more interested in the shape of orbits, that is, 
in $r$ as a function of $\theta$, than in their time history. The 
angular momentum relation can then be used directly to convert 
(12) into the differential equation for the orbit; this gives 
\begin{equation} 
\frac{(mc^2+E)^2}{c^2l^2g_{00}}-\frac{g_{rr}}{r^4} 
\biggl(\frac{dr}{d\theta}\biggr)^2-\frac{1}{r^2}= 
\frac{m^2c^2}{l^2}. 
\end{equation} 
The solution may thus be determined by a quadrature: 
\begin{equation} 
\triangle\theta=\int\biggl[\frac{(mc^2+E)^2}{c^2l^2g_{00}}- 
\frac{m^2c^2}{l^2}-\frac{1}{r^2}\biggr]^{-1/2} 
\frac{g_{rr}^{1/2}dr}{r^2}. 
\end{equation} 

At perihelia and aphelia, $r$ reaches its minimum and maximum 
values $r_-$ and $r_+$, and at both points $dr/d\theta$ vanishes, 
so (13) gives 
\begin{equation} 
\frac{(mc^2+E)^2}{c^2l^2g_{00}(r_{\pm})}-\frac{1}{r_{\pm}^2}= 
\frac{m^2c^2}{l^2}, 
\end{equation} 
where $g_{00}(r_{\pm})=1-2GM/c^2r_{\pm}$. From these two 
equations we can derive values for the two constants of the motion: 
\begin{equation} 
\biggl(1+\frac{E}{mc^2}\biggr)^2= 
\frac{r_+^2-r_-^2}{r_+^2g_{00}^{-1}(r_+)-r_-^2g_{00}^{-1}(r_-)}, 
\quad \frac{m^2c^2}{l^2}= 
\frac{r_+^{-2}g_{00}(r_+)-r_-^{-2}g_{00}(r_-)}{g_{00}(r_-)-g_{00}(r_+)}. 
\end{equation} 
The expressions for the energy and angular momentum appear 
here in somewhat different forms involving the metric tensors 
$g_{00}(r_{\pm})$, but their equivalence in the limit as 
$g_{00}\to 1$ with the respective nonrelativistic Newtonian 
relations are shown by expanding the equations to a first 
approximation: 
\begin{equation} 
E\simeq-\frac{GMm}{r_++r_-}, \quad 
l^2\simeq\frac{2GMm^2}{r_+^{-1}+r_-^{-1}}. 
\end{equation} 
Using the exact values of the constants given by (16) in (14) 
yield the formula for $\triangle\theta$ as 
\begin{equation} 
\triangle\theta=\int\biggl[\frac{r_+^{-2}(g_{00}^{-1}(r)- 
g_{00}^{-1}(r_-)) - r_-^{-2}(g_{00}^{-1}(r)-g_{00}^{-1}(r_+))} 
{g_{00}^{-1}(r_+)-g_{00}^{-1}(r_-)}-\frac{1}{r^2}\biggr]^{-1/2} 
\frac{g_{rr}^{1/2}(r)dr}{r^2}. 
\end{equation} 
We can make the argument of the first square root in the integrand a 
quadratic function of $1/r$ which vanishes at $r=r_{\pm}$, so 
\begin{equation} 
\triangle\theta\simeq\int\biggl[C\biggl(\frac{1}{r_-}- 
\frac 1r\biggr)\biggl(\frac 1r -\frac{1}{r_+}\biggr)\biggr]^{-1/2} 
\biggl(1+\frac{GM}{c^2r}\biggr)\frac{dr}{r^2}, 
\end{equation} 
where $C\simeq 1-(2GM/c^2)(r_+^{-1}+r_-^{-1})$. The constant $C$ 
could be determined by letting $r\to\infty$. 

We can obtain the same result much more simply. It is both easier 
and more instructive to expand $g_{00}$ in the formal solution (14). 
It preserves the advantage that the orbit equation is evaluated 
in terms of the energy and the angular momentum of the system. 
Note that we have to expand to second order in $GM/c^2r$. 
The angle swept out by the position vector is then given by (14) as 
\begin{equation} 
\triangle\theta\simeq\int\biggl[ 
\frac{2mE}{l^2}\biggl(1+\frac{E}{2mc^2}\biggr)+ 
\frac{2GMm^2}{l^2r}\biggl(1+\frac{2E}{mc^2}\biggr)- 
\frac{1}{r^2}\biggl(1-\frac{4G^2M^2m^2}{c^2l^2}\biggr) 
\biggr]^{-1/2}\biggl(1+\frac{GM}{c^2r}\biggr)\frac{dr}{r^2}. 
\end{equation} 
As it stands, this integral is of the standard form. The integrand differs 
from the corresponding nonrelativistic expression in that the second 
term in each pair of parenthesis represents the relativistic correction. 
In form it is the general-relativity analogue of Sommerfeld's 
treatment of the hydrogen atom in special relativity. It has been said 
that we need $g_{00}$ to second order in $GM/c^2r$ to calculate 
$\triangle\theta$ to first order. To put this another way, the high 
accuracy of the orbit precession serves as a touchstone for the 
possible forms of $g_{00}$ by requiring the degree of agreement 
to second order. The procedure described here is particularly 
simple and is sufficient to enable one to confirm the fact. 

On carrying out the integration, the equation of the orbit is found 
to be 
\begin{equation} 
\frac 1r\simeq A\biggl[1+\epsilon\cos\biggl( 
(\theta-\theta_0)\biggl(1-\frac{3G^2M^2m^2}{c^2l^2}\biggr) 
-\frac{GM}{c^2r^2}\biggl(\frac{dr}{d\theta}\biggr) 
\biggr)\biggr], 
\end{equation} 
where 
\[ A\simeq \frac{GMm^2}{l^2} 
\biggl(1+\frac{2E}{mc^2}+\frac{4G^2M^2m^2}{c^2l^2}\biggr), 
\quad \epsilon\simeq\biggl[1+\frac{2El^2}{G^2M^2m^3} 
\biggl(1-\frac{7E}{2mc^2}-\frac{4G^2M^2m^2}{c^2l^2}\biggr) 
\biggr]^{1/2}, \] 
and $\theta_0$ is a constant of integration. In addition to the motion 
of a planet's perihelion of $2\pi(3G^2M^2m^2/c^2l^2)$ per revolution, 
the relativity effect produces the term $(GM/c^2r^2)(dr/d\theta)$ 
in the angle swept out by the radius vector of the planet. This 
term is not a new result but merely a result of rewriting the square 
root in the integrand which the integration of (20) actually yields, 
using (14) to a first approximation. It is evident therefore that the 
relativity effect in planetary motion obeying (19) or (20) is to 
cause not only the precession of the perihelion of the orbit 
of a planet but also the change in the angular displacement of 
the planet due to its radial velocity. The additional change 
appearing in the angular displacement of the planet, which 
does not appear in a circular orbit, might be an effect due to 
the finite velocity of propagation of the solar gravitational field. 

The formulation presented in this paper is mathematically 
equivalent to the familiar formulations. There are therefore no 
fundamentally new results. However, the point of view which 
has been taken here regarding the central force problem differs 
from the usual point of view. There is a certain lack of energy 
concept in the geometrical approach to the subject. 
A prominent feature of the present formulation is that 
the customary concepts of classical mechanics are 
emphasized throughout within the mathematical framework 
required by general relativity. This is very helpful for grasping 
the physical meaning behind the mathematical development. 
The present point of view offers a distinct advantage.

\end{document}